# New Eruptive YSOs from SPICY and WISE

**Carlos Contreras Peña** [1,2,*], **Mizna Ashraf**[3], **Jeong-Eun Lee** [1,4], **Gregory Herczeg** [5,6], **Phil Lucas** [7], **Zhen Guo** [8,9,10,7], **Doug Johnstone** [11,12], **Ho-Gyu Lee** [13], **and Jessy Jose** [3]

[1]Department of Physics and Astronomy, Seoul National University, Seoul 08826, Republic of Korea
[2]Research Institute of Basic Sciences, Seoul National University, Seoul 08826, Republic of Korea
[3]Indian Institute of Science Education and Research (IISER) Tirupati, Tirupati 517 507, India
[4]SNU Astronomy Research Center, Seoul National University, Seoul 08826, Republic of Korea
[5]Kavli Institute for Astronomy and Astrophysics, Peking University, Beijing 100871, People's Republic of China
[6]Department of Astronomy, Peking University, Beijing 100871, People's Republic of China
[7]Centre for Astrophysics Research, University of Hertfordshire, Hatfield, AL10 9AB, UK
[8]Instituto de Física y Astronomía, Universidad de Valparaíso, Casilla 5030, Valparaíso, Chile
[9]Núcleo Milenio de Formación Planetaria (NPF), Casilla 5030, Valparaíso, Chile
[10]Departamento de Física, Universidad Tecnicá Federico Santa María, Avenida España 1680, Valparaíso, Chile
[11]NRC Herzberg Astronomy and Astrophysics, Victoria, BC, V9E 2E7, Canada
[12]Department of Physics and Astronomy, University of Victoria, Victoria, BC, V8P 5C2, Canada
[13]Korea Astronomy and Space Science Institute, Daejeon 34055, Republic of Korea

*Corresponding Author: C. Contreras Peña, ccontreras@snu.ac.kr

## Abstract

This work presents four high-amplitude variable YSOs ($\simeq 3$ mag at near- or mid-IR wavelengths) arising from the SPICY catalog. Three outbursts show a duration that is longer than 1 year, and are still ongoing. And additional YSO brightened over the last two epochs of NEOWISE observations and the duration of the outburst is thus unclear. Analysis of the spectra of the four sources confirms them as new members of the eruptive variable class. We find two YSOs that can be firmly classified as bona fide FUors and one object that falls in the V1647 Ori-like class. Given the uncertainty in the duration of its outburst, an additional YSO can only be classified as a candidate FUor. Continued monitoring and follow-up of these particular sources is important to better understand the accretion process of YSOs.

## 1. Introduction

The observed spread in luminosities of Class I YSOs in star-forming regions contradicts the expectations from standard models of star formation (Kenyon et al. 1990; Evans et al. 2009). Episodic accretion has been invoked to solve this issue. In this model, stars would gain most of their mass in short-lived episodes of high accretion followed by long periods of quiescent low-level accretion (see e.g. Fischer et al. 2022). The long time spent at high accretion rates can have long-term effects on the structure of the central star (Baraffe et al. 2017). The outbursts can alter the chemistry of protoplanetary discs (Artur de la Villarmois et al. 2019), the location of the snowline of various ices (Cieza et al. 2016; Lee et al. 2019), aid in the formation of planetary systems similar to the solar system (Hubbard 2017) and could have an effect on the orbital evolution of planets, if present (Boss 2013; Becker et al. 2021).

Young stellar objects (YSOs) display sudden rises in brightness that provide direct observational evidence for stars undergoing episodic accretion (Hartmann & Kenyon 1996). These YSOs are classified according to the photometric and spectroscopic characteristics of the outburst (most commonly known as FUor and EX Lupi-type outbursts, Fischer et al. 2022). Eruptive variable YSOs are still a rare class of variable stars, with no more than 40 objects classified as such over the last 85 years (Connelley & Reipurth 2018; Hillenbrand et al. 2018), although the number has been increasing thanks to continuous monitoring across optical, near-IR, mid-IR and sub-mm wavelengths (see e.g. Contreras Peña et al. 2017; Guo et al. 2021; Lee et al. 2021; Park et al. 2021; Contreras Peña et al. 2023).

There are some caveats that affect any discussion on the impact of outbursts in YSO evolution. There is still uncertainty in the frequency of these events (Fischer et al. 2022) and whether all YSOs go through these episodes of high accretion. Millimetre continuum observations with ALMA have shown that FUor disks are more massive and compact than the disks of other classes of eruptive variables (Cieza et al. 2018) and those of regular Class II and Class I YSOs (Kóspál et al. 2021). This could imply that not all YSOs gain their mass through episodic accretion, but FUors, or other classes of eruptive variable stars, are instead a type of YSOs that follow a particular path in their evolution that leads to the episodes of high accretion (Fischer et al. 2022).

Increasing the sample of YSOs that undergo large accretion-related outbursts is a key step to understanding the universality of episodic accretion, and the frequency of these events among YSOs. In addition, continuous multi-wavelength monitoring of outbursting YSOs can yield insights into the physical mechanisms triggering the outburst (Cleaver et al. 2023).

As part of an ongoing effort to detect and characterize eruptive YSOs, we searched for high amplitude variability in sources from the SPICY catalogue (Kuhn et al. 2021), using multi-epoch WISE/NEOWISE mid-IR observations. Spectroscopic follow-up of 11 sources with amplitudes larger than 1.3 mag allows us to confirm four YSOs as new members of the eruptive variable class. In this paper, we show the data for the 11 sources but present a more detailed discussion on the four sources that can be confirmed as new members of the eruptive variable class.

## 2. Observations

The YSO sample arises from The *Spitzer*/IRAC Candidate YSO catalog for the inner Galactic midplane (SPICY). Kuhn et al. (2021) use a random forest classification to select YSOs from *Spitzer* photometry obtained during the cryogenic mission. This includes seven *Spitzer*/IRAC surveys covering 613 square degrees. The data is also augmented with near-infrared surveys 2MASS (Skrutskie et al. 2006), UKIDSS Galactic Plane Survey (GPS, Lawrence et al. 2007; Lucas et al. 2008) and VVV (Minniti et al. 2010; Saito et al. 2012). Additional information, such as spatial distribution and variability, is used to corroborate the nature of YSOs for objects in the catalog. After applying these criteria, the SPICY catalog contains 117446 YSOs.

### 2.1. Photometry

This work uses mid-IR photometry from all-sky observations of the *WISE* space telescope. *WISE* surveyed the entire sky in four bands, $W1$ (3.4 $\mu$m), $W2$ (4.6 $\mu$m), W3 (12 $\mu$m), and W4 (22 $\mu$m), with the spatial resolutions of 6.1″, 6.4″, 6.5″, and 12″, respectively, from January to September of 2010 (Wright et al. 2010). The survey continued as the NEOWISE Post-Cryogenic Mission, using only the $W1$ and $W2$ bands, for an additional 4 months (Mainzer et al. 2011). In September 2013, *WISE* was reactivated as the NEOWISE-reactivation mission (NEOWISE-R, Mainzer et al. 2014). NEOWISE-R is still operating, and the latest released data set contains observations until mid-December 2022. For each visit to a particular area of the sky, *WISE* performs several photometric observations for ∼few days. Each area of the sky is observed in a similar way every ∼6 months.

For the analysis of SPICY YSOs, we used all the available data from the *WISE* telescope for observations between 2010 and 2022. The single-epoch data was collected from the NASA/IPAC Infrared Science Archive (IRSA) catalogues using a 3″ radius from the coordinates of the YSO. For each source, we averaged the single epoch data taken a few days apart to produce 1 epoch of photometry every 6 months (following the procedures described in Park et al. 2021). The majority of sources have between 19 and 20 epochs of mid-IR photometry.

To search for candidate YSOs where large changes in the accretion rate are driving variability, we selected sources that had 14 or more epochs in both $W1$ and $W2$ filters and that fulfilled $\Delta W1 \geq 1.3$ and $\Delta W2 \geq 1.3$ mag. These large amplitudes are expected for accretion-driven outbursts (Scholz et al. 2013; Liu et al. 2022; Hillenbrand & Rodriguez 2022). The cut in the number of epochs and amplitude yields 1202 sources. Visual inspection of the light curves allowed us to select a sample of YSOs for spectroscopic follow-up.

For all of the YSOs presented in this work, we also collected additional data from various publicly available catalogues. These included the *Spitzer* surveys (data which is included in the SPICY catalogue), 2MASS (Cutri et al. 2003), and the UKIDSS Galactic Plane Survey (Lucas et al. 2008, 2017). The magnitudes in channels 1 and 2 from *Spitzer*/IRAC are converted into the *WISE* system using the equations from Antoniucci et al. (2014). We note that the *Spitzer* data is presented as it is a useful indicator of large amplitudes changes (or lack of thereof) between ∼2004 and 2010. Given the uncertainties in the process of transforming between different photometric systems, we do not attempt to draw any conclusions based on any apparent color changes between *Spitzer* and *WISE* observations.

### 2.2. Spectroscopy

We obtained near-IR spectra of SPICY YSOs 79425, 95397, 97855, 99341, 100587, 103300, 104367, 109331 and 115884 on 11–12 July 2023 (HST) with SpeX (Rayner et al. 2003) mounted at the NASA Infrared Telescope Facility (IRTF) on Mauna Kea (programme 2023A974, PI Ashraf). The cross-dispersed spectra cover 2.1–2.5 $\mu$m spectra at $R \sim 2000$, obtained with the 0.5″slit. Total integration times ranged from 480 to 1920 s, with individual exposures of 60 to 120 s. Bright A0V standard stars were observed for telluric calibration. All spectra were reduced and calibrated using Spextool version 4.1 (Cushing et al. 2004).

From the nine objects observed with IRTF/SpeX, only in four cases we can confirm a classification as eruptive variable YSOs. The mid-IR light curves and spectroscopic characteristics of the remaining five YSOs are presented in Appendix A.

**Table 1.** YSO Sample

| YSO ID | Other Name | $\alpha$ (J2000) | $\delta$ (J2000) | Class | Distance (kpc) | $\Delta K$ | $\Delta W1$ | $\Delta W2$ | Spectral Class | Photometric Class | Final Class |
|---|---|---|---|---|---|---|---|---|---|---|---|
| SPICY 97855 | — | 19:05:26.31 | +05:57:34.87 | FS | — | — | 3.16 | 2.77 | FUor | candidate FUor | candidate FUor* |
| SPICY 99341 | SSTOERC G043.2810−00.3252[a] | 19:11:38.79 | +09:02:59.11 | II | 3 | 4 | 3.03 | 2.84 | FUor | FUor | bona-fide FUor |
| SPICY 100587 | 2MASS J19171791+1116323[b] | 19:17:17.93 | +11:16:32.29 | II | — | 2.6 | 1.47 | 1.66 | FUor | FUor | bona-fide FUor |
| SPICY 109331 | [KMH2014] J202432.54+374949.21[c] | 20:24:32.55 | +37:49:49.22 | I | 3.84 | — | 3.29 | 2.43 | EX Lupi | V1647 Ori? | V1647 Ori |

[a] classified as a YSO in the W49 SFR by Saral et al. (2015), [b] classified as a candidate YSO by Robitaille et al. (2008), [c] classified as a protostellar candidate in Cygnus-X by Kryukova et al. (2014), * The final classification is uncertain as the outburst is still ongoing.

**Table 2.** Equivalent widths (in units of Å) for the detected lines in the near-IR spectra of SPICY YSOs.

| YSO ID | $H_2$ (2.12 $\mu$m) | Br$\gamma$ (2.16 $\mu$m) | Na I (2.206, 2.209 $\mu$m) | Ca I (2.263, 2.266 $\mu$m) | $^{12}CO_{\nu=2-0}$ (2.29 $\mu$m) |
|---|---|---|---|---|---|
| SPICY 97855 | — | — | $1.4 \pm 0.5$ | $0.2 \pm 0.5$ | $14.4 \pm 0.9$ |
| SPICY 99341 | — | — | $1.5 \pm 0.5$ | $1.1 \pm 0.5$ | $14.8 \pm 0.9$ |
| SPICY 100587 | — | $-2.3 \pm 0.4$ | $1.8 \pm 0.5$ | $1.2 \pm 0.5$ | $20.5 \pm 1.1$ |
| SPICY 109331 | $-2.7 \pm 0.4$ | $-5.8 \pm 0.5$ | — | — | $-10.2 \pm 0.9$ |

From this point onward, we focus on the four YSOs that show spectroscopic characteristics of the eruptive variable class. Table 1 shows the information for the four SPICY YSOs.

For the four SPICY YSOs we estimated the equivalent widths (EWs) by integrating the continuum normalized fluxes in the regions of five spectral features that are commonly detected in the spectra of YSOs ($H_2$, Br$\gamma$, Na I, Ca I and $^{12}$CO). For each line, we estimated the value of EW after introducing Gaussian noise to the flux. This step is repeated 1000 times for each line. The final value and its error are estimated as the mean and standard deviation over the 1000 measurements. The values of EW for the detected lines in each spectrum are presented in Table 2.

## 3. New Eruptive YSOs

This section discusses the various characteristics of individual sources, including information from the literature, photometric behavior, and observed spectroscopic characteristics.

***SPICY 97855***: The object is classified as a flat-spectrum YSO in the SPICY catalogue (Kuhn et al. 2021). There is little previous information about the source. It was observed as part of the UKIDSS GPS survey, but it showed no variability over the two epochs (taken 4 years apart), with $K \sim 17.2$ mag in both observations. Figure 1 shows that the source became brighter in the 2022 NEOWISE observations, where it increased by 3.16 and 2.77 mags in $W1$ and $W2$, respectively.

In the initial stages, the color of the source in the $W1$ versus $W1-W2$ color-magnitude diagram (CMD, Figure 1) becomes bluer as the YSO becomes brighter in $W1$. This type of evolution agrees with changes driven by accretion outbursts (Antoniucci et al. 2014). Although this change also follows the expected variability from changes in the extinction along the line of sight, the large amplitudes argue against this mechanism as the main driver of the variability (see also Section 5 in Contreras Peña et al. 2023). Interestingly, the YSO becomes redder during the large brightening over the most recent epochs of NEOWISE observations. This is similar to the outbursts in HOPS267, LDN 1455 IRS3 (Contreras Peña et al. 2023), EC53 (Lee et al. 2020), and WISEA J142238.82−611553.7 (Lucas et al. 2020). In the case of EC53, Lee et al. (2020) argue that mass builds up at longer radii in the disk, where at some point this bloackage becomes unstable, draining into the inner regions of the disk and then the star. Right before the outburst, the system becomes redder due the the building up of material leading to changes in the scale height of the inner disk and increasing extinction towards the source. The colour behaviour in the aforementioned sources may point to a similar mechanism driving the outburst as in EC53.

The light curve and mid-IR CMD point to a gradual increase in the mid-IR brightness of the source. This is similar to the behavior of the mid-IR light curve of Gaia 17bpi (Hillenbrand et al. 2018). Although we lack information at optical wavelengths, this might indicate that we are observing an outside-in type of outburst in SPICY97855, where the outburst begins in a cooler region of the disk and then propagates inwards leading to a later occurrence of the optical outburst (Fischer et al. 2022).

The IRTF spectra taken at two different nights (Figure 1) show strong $^{12}$CO absorption beyond 2.29 $\mu$m. There is also some weak detection of Na I and Ca I absorption, and there appear to be some differences in the continuum level. However, the signal-to-noise ratio (SNT) varies between 15–20 over the two nights. The low SNR does not allow us to infer any conclusions on possible differences between the two nights, as these may arise from the uncertainties in the process of correcting of telluric features.

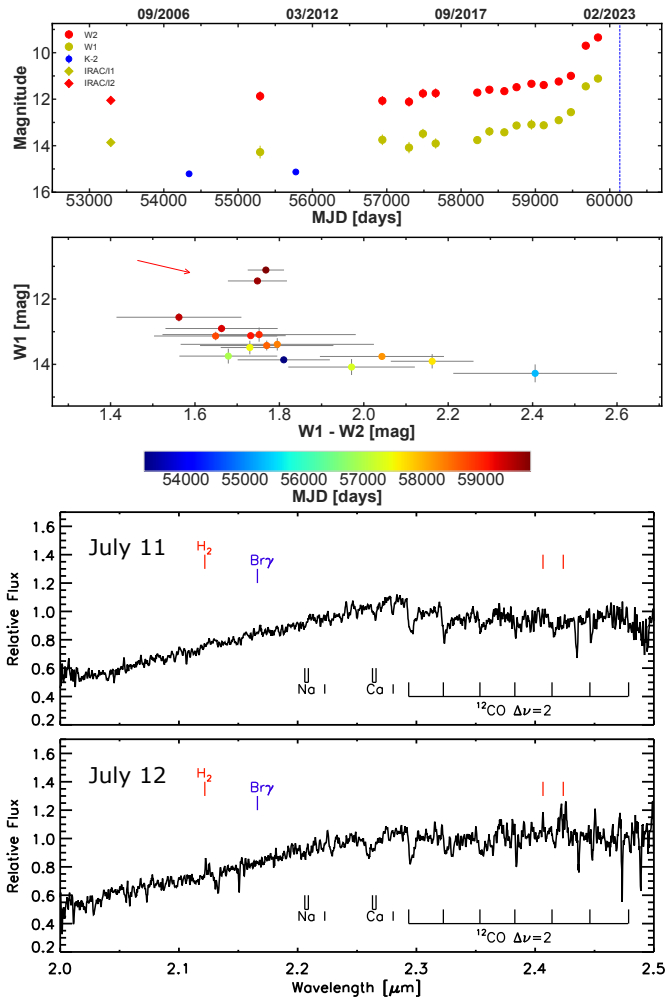

**Figure 1.** (top) K-band (shifted by $-2$ magnitudes for this YSO, blue), *WISE* $W1$ and *Spitzer*/IRAC channel 1 (yellow circles and diamonds), and *WISE* $W2$ and *Spitzer*/IRAC channel 2 (red circles and diamonds) light curve of SPICY 97855. The magnitudes from channels 1 and 2 from *Spitzer*/IRAC were converted into the WISE photometric system using the equations from Antoniucci et al. (2014). A dashed vertical blue line marks the date of spectroscopic observations. (middle) Mid-IR color-magnitude diagram for SPICY 97855 where the color of the data points indicates the observations dates (MJD), as shown in the color bar. The CMD includes the data *Spitzer*/IRAC channel 1 and 2 converted into the WISE photometric system. The red arrow marks the reddening line for $A_V = 20$ mag, using the extinction law of Wang & Chen (2019). (bottom) IRTF/SpeX spectrum of SPICY 97855. The location of typical emission/absorption features in YSOs are indicated in the figures. These include Na I, Ca I and $^{12}$CO (black), Br$\gamma$ (blue) and different transitions of H$_2$ (red lines).

*SPICY 99341*: This is a Class II YSO in the SPICY catalogue (Kuhn et al. 2021). It was previously classified as a YSO in an infrared study of the W49 massive star-forming region (Saral et al. 2015). Lucas et al. (2017) include it in their sample of high-amplitude variables arising from the UKIDSS GPS survey (source 266, $\Delta K = 4$ mag). Lucas et al. (2017) provide a distance of 3 kpc to the source due to its possible association with the molecular cloud GRSMC 43.30−0.33 (Simon et al. 2001). The latter agrees, within the errors, with the distance to the source, estimated from *Gaia* observations, of $2.9^{+1.2}_{-1.1}$ kpc

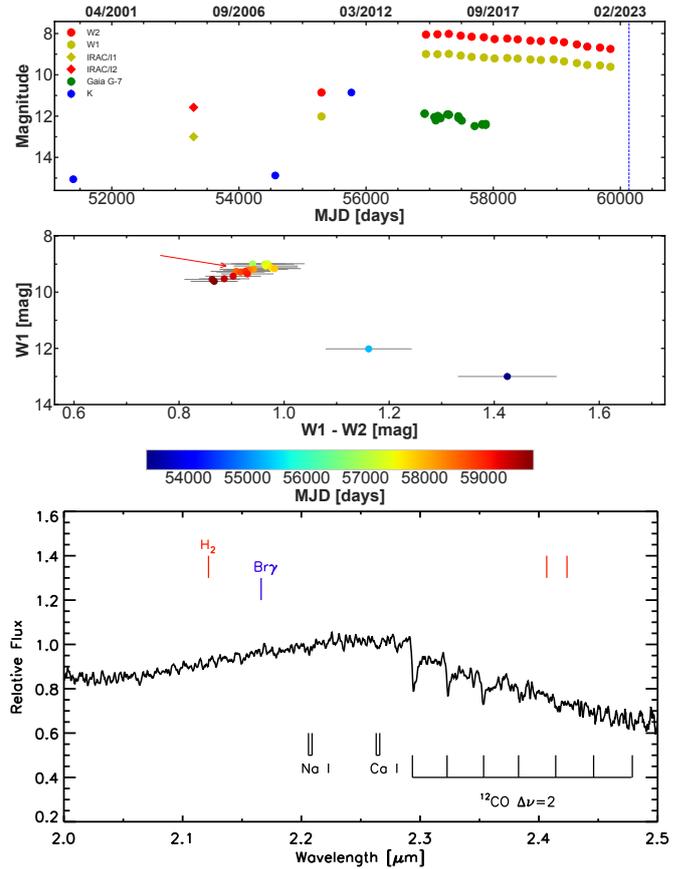

**Figure 2.** (top) Light curve of SPICY 99341. Here we include photometric data from *Gaia* (green circles), which is shifted by $-7$ magnitudes. (middle) $W1$ versus $W1 - W2$ color-magnitude diagram, and (bottom) IRTF/SpeX spectrum of SPICY 99341. Symbols and labels are the same as in Figure 1.

(Bailer-Jones et al. 2021).

The light curve of Figure 2 shows that the source appears relatively stable between 2MASS and the first epoch of UKIDSS GPS observations. The high-amplitude change at the second GPS epoch (Lucas et al. 2017) is due to the outburst, which occurred after the WISE observations between 2010–2011. The source is optically visible and is detected by *Gaia* at $G \simeq 19$ (Gaia Collaboration 2022), with the source showing a long-term decline, similar to the observed behavior at mid-IR. Given the $W1 - G$ colour of $\simeq 10$ mag in the first epoch of NEOWISE observations, and assuming a similar colour before the outburst, puts the (pre-outburst) brightness of the source at $G \simeq 22$ mag.

The mid-IR CMD (Figure 2) shows that the YSO becomes much bluer, changing from $(W1 - W2) \simeq 1.4$ mag during quiescence, and reaching $(W1 - W2) \simeq 0.9$ mag as the source gets to the maximum point in the light curve. The trend seems to follow the reddening line, but similarly to SPICY 97855, the large amplitude is not expected to be driven by changes in the extinction along the line of sight.

The IRTF spectrum (Figure 2) is dominated by strong $^{12}$CO absorption at 2.29 $\mu$m, with some weaker detection at the wavelengths of Ca I and Na I. The spectrum was not observed at the peak brightness of the source, with the source being

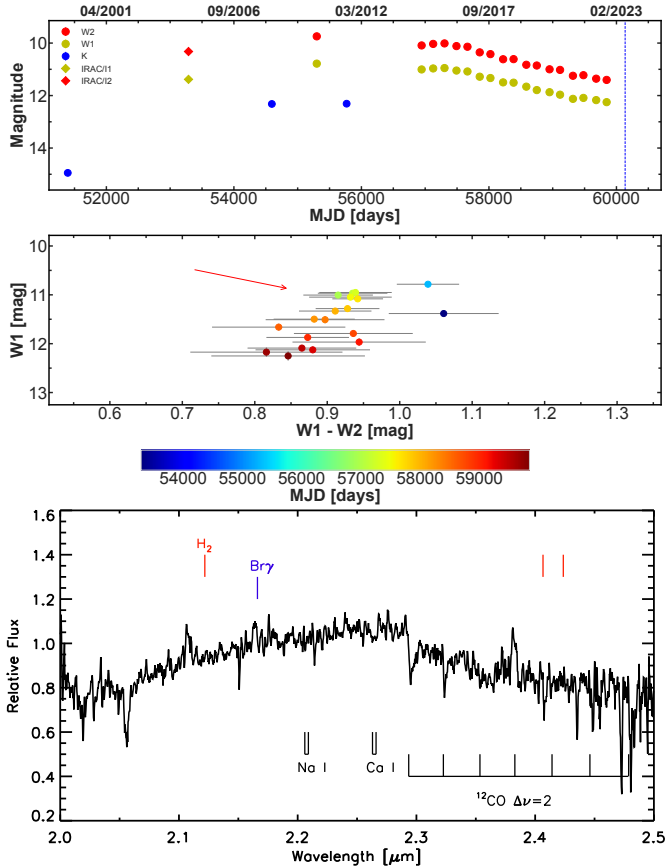

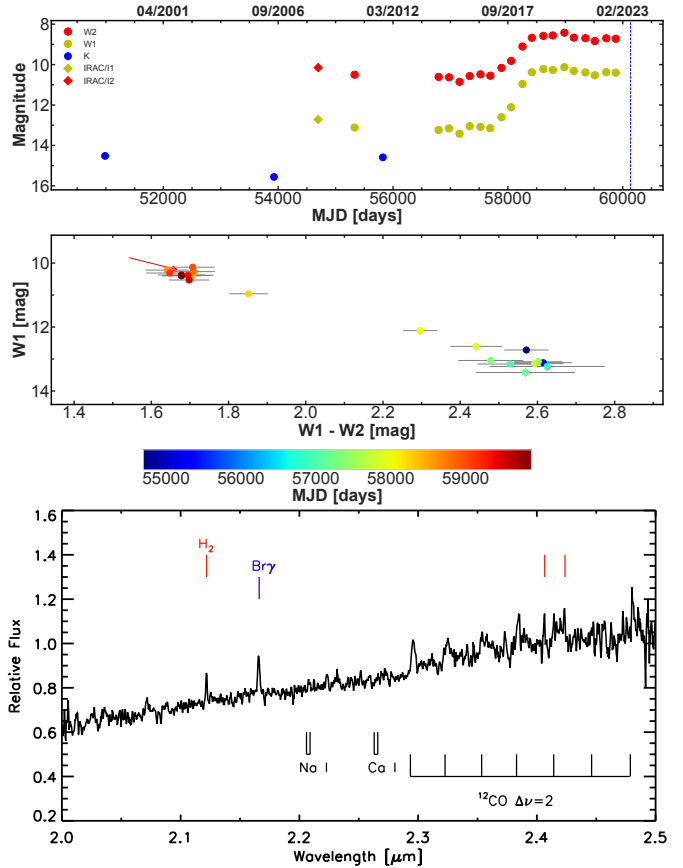

**Figure 3.** (top) Light curve, (middle) $W1$ versus $W1-W2$ color-magnitude diagram, and (bottom) IRTF/SpeX spectrum of SPICY 100587. Symbols and labels are the same as in Figure 1. The broad features at 2.05, 2.11 and 2.38 $\mu$m are artifacts arising from the data reduction process.

**Figure 4.** (top) Light curve, (middle) $W1$ versus $W1-W2$ color-magnitude diagram, and (bottom) IRTF/SpeX spectrum of SPICY 109331. Symbols and labels are the same as in Figure 1.

$\sim$1 mag fainter than the first NEOWISE epoch. However, the YSO is likely still at a high accretion state during the IRTF observations.

*SPICY 100587*: This is another Class II YSO in the SPICY catalogue (Kuhn et al. 2021). It was previously classified as a candidate YSO due to its intrinsically red near- to mid-IR colors (Robitaille et al. 2008). The object was observed as part of the UKIDSS GPS survey but it shows a constant brightness at $K=12.3$ mag over the two epochs. No distance information could be found in the literature for this source.

The mid-IR amplitude is the smallest among the four SPICY YSOs presented in this work (Table 1). However, in K-band, the brightness jumped about $\sim$3 mag between 1998 and 2008. On the other hand, in mid-IR, the brightness is relatively constant between *Spitzer/IRAC* data (2004) and the first epoch of *WISE* observations in 2010. Thus, the outburst must have occurred between 1998 and 2004.

The mid-IR color of this source during NEOWISE observations appears to be relatively stable as the source fades. However, there is some indication of the YSO becoming bluer as it becomes fainter. This trend would show that the $W2$ brightness of the source is fading faster than its brightness at $W1$. Although the trend is only weakly seen, it might imply that the cooling front of the instability that leads to the outburst (e.g., a thermal instability model Bell & Lin 1994) is moving inwards.

The IRTF spectrum (Figure 3) is similar to that of SPCIY 97855 and SPICY 99341, showing strong $^{12}$CO absorption at 2.29 $\mu$m. We also detect the presence of Na I and Ca I absorption, as well as possible weak Br$\gamma$ emission.

*SPICY 109331*: This is the only YSO in our sample that is classified as a Class I YSO, based on the spectral index, $\alpha=1.1$, estimated from its $3.4-22$ $\mu$m colour (Kuhn et al. 2021). It was previously given a similar classification in Kryukova et al. (2014), as part of an infrared study of objects towards Cygnus-X. Based on *Spitzer* IRAC and MIPS observations, Kryukova et al. (2014) estimate a luminosity of $L=13.2$ $L_\odot$ for the source. The object was also observed as part of the UKIDSS GPS survey, but it shows a $\sim$1 mag variability between two epochs measured at $K=14.6$ mag and $K=15.7$ mag, taken 5 years apart. The YSO is also the candidate driving source of a H$_2$ outflow (OF240, Makin & Froebrich 2018). Finally, the source is located at $2.6''$ from a Hi-Gal compact source at a distance of 3.84 kpc (Mège et al. 2021).

Prior to the outburst, the source shows a relatively stable brightness, with some near-IR variability that could be attributed to a different physical mechanism, unrelated to the accretion-driven outburst. NEOWISE data shows that the outburst started at MJD $\sim$ 58000 and is still ongoing

(see Figure 4).

The mid-IR CMD (Figure 4) shows a similar trend to that of SPICY 99341, and this source becomes bluer by $\Delta(W1 - W2) \simeq 0.9$ mag as it becomes brighter. This agrees with the expectations of an accretion-driven outburst.

The IRTF spectrum (Figure 4) is different from the other sources presented in this work, showing $^{12}$CO, Br$\gamma$ and H$_2$ emission. No emission of Na I or Ca I could be detected.

## 4. Discussion

Traditionally, eruptive YSOs are classified into two sub-categories, FUors and EX Lupi-type.

FUors show high-amplitude outbursts ($\Delta V \sim 6$ mag) that can last for decades (typically longer than 10 years) due to the sudden increase of the accretion rate, which can reach as high as $10^{-4} M_\odot$ yr$^{-1}$ (Hartmann & Kenyon 1996). During outbursts, FUors are characterized by an absorption spectrum with very few emission lines. In the near-IR, $^{12}$CO bandhead absorption (2.29 $\mu$m) and a triangular H-band continuum due to H$_2$O absorption (1.33 $\mu$m) are generally seen. During the outburst, the accretion disk dominates emission in the system, where absorption lines arise due to a cooler disk surface compared with the viscously-heated midplane (Herbig 1977; Hartmann & Kenyon 1996; Reipurth & Aspin 2010; Connelley & Reipurth 2018; Liu et al. 2022).

EX Lup outbursts (originally named EXors in Herbig 1989), are considered as the less dramatic counterparts of FUors. The outbursts in these systems can last a few weeks to several months and reach similar optical amplitudes to FUor outbursts. In addition, outbursts in these systems have sometimes been seen to be repetitive (Cruz-Sáenz de Miera et al. 2023; Wang et al. 2023). The spectra of EX Lup type objects are dominated by emission lines during maximum light. In the near-IR, Na I (2.206 $\mu$m) and $^{12}$CO bandhead emission arise from the surface layers of a hot inner disk. These features go into absorption during the quiescent state (Lorenzetti et al. 2007, 2009). In general, EX Lup type objects observed at photometric minima were found to be no different than typical T Tauri stars (Herbig 1989, 2008; Lorenzetti et al. 2012; Audard et al. 2014).

More recent discoveries of outbursts, however, have blurred this classification. The majority of new discoveries tend to show a mixture of spectroscopic and photometric characteristics between those of EX Lup and FUor outbursts. For example, in the sample of outbursts from the VVV survey, the majority of long-duration (longer than 1900 days) outbursts show emission line spectra (Guo et al. 2021). Given these mixed characteristics, the new discoveries have been designated as either V1647 Ori-like, MNor, or Peculiar outbursts (Contreras Peña et al. 2017; Connelley & Reipurth 2018). For a discussion on the classification issue, see also Fischer et al. (2022); Contreras Peña et al. (2023).

The four SPICY YSOs described in this work show long-term (longer than 1 year), high amplitude ($\Delta > 2.5$ mag) outbursts, as well as the appearance of strong $^{12}$CO absorption/emission in their spectra. These characteristics are all consistent with those of objects undergoing outbursts due to large changes in the accretion rate. In the following, we attempt to place the characteristics of these outbursts into the categories of eruptive variable YSOs described above.

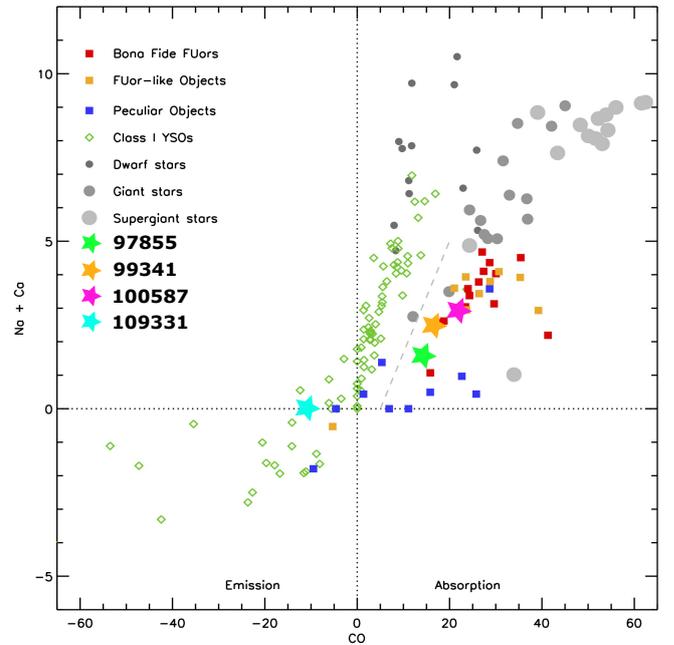

**Figure 5.** Equivalent width of Na I and Ca I versus $^{12}$CO for a sample of Class I YSOs, field dwarfs and giant stars, and eruptive YSOs (Figure 9 in Connelley & Reipurth 2018). The values for the four SPICY YSOs are presented as star symbols.

The spectra of the SPICY YSOs 97855, 99341, and 100587 strongly resemble that of FUor outbursts (Connelley & Reipurth 2018). The comparison of EWs from Na I+Ca I versus that of $^{12}$CO (Figure 5) shows that these three YSOs are located in the region of bona fide FUors presented by Connelley & Reipurth (2018).

The high-amplitude of the brightness change of the three YSOs also agrees with the FUor of classification. SPICY 97855 and 99341 show higher mid-IR amplitude than SPICY 100587, but this is due to WISE observations covering the outbursts in the former cases, while the mid-IR variability of SPICY 100587 shows that the object is slowly returning to quiescence.

It is unclear how long the outburst of SPICY 97855 will last. The current data suggest that the outburst is still ongoing, making it a $\sim$2 year outburst. The intermediate duration of the outburst would place it, currently, in the V1647 Ori-like class. However, it is still too early to discard an FUor classification. Given this uncertainty, SPICY 97855 can only be classified as a candidate FUor. In the case of SPICY 99341, the outburst probably started around 2010–2011, which indicates a duration of 10 years or longer. A similar conclusion can be reached for SPICY 100587 as the outburst started before *Spitzer* observations and is still ongoing. Therefore, we can firmly place SPICY 99341 and 100587 as new additions to the bona-fide FUor class. Due to the uncertainty in the duration of the outburst, this is less clear for SPICY 97855, and we can temporarily place it as a candidate FUor.

SPICY 109331 is the only object in our sample that displays an emission line spectrum. This type of spectrum during outburst is more consistent with the EX Lup type classification (Lorenzetti et al. 2012). The EW of Na I+Ca I versus $^{12}$CO of the spectrum puts in the region where other peculiar (as defined by Connelley & Reipurth 2018) outbursts are located (Figure 5).

The outburst in this source is still ongoing, and the current data shows an outburst duration of ∼5 years. This is longer than the expectation for EX Lup type objects. Therefore, this object is classified as V1647 Ori-like.

SPICY 109331 is an interesting source as it shows a large infrared luminosity of $L = 13.2\ L_\odot$ before the outburst. This luminosity is towards the higher end of the luminosity of low-mass protostars (Dunham et al. 2014). This could imply a larger mass for the YSO, which could explain the observed spectrum during the outburst, despite its high amplitude. According to Liu et al. (2022), larger accretion rates need to be reached for YSOs with higher masses (in the range 2–3 $M_\odot$) for the viscously-heated disk to dominate emission in the system and therefore show the absorption spectrum typical of FUors.

## 5. Summary

To characterize episodic accretion during the evolution of young stellar objects, we have been monitoring the mid-IR variability of large samples of known YSOs (Park et al. 2021; Contreras Peña et al. 2023). Using the latest data release from NEOWISE (Mainzer et al. 2014) we searched for a sample YSOs in the SPICY catalogue Kuhn et al. (2021) that show high-amplitude ($\Delta > 1.3$ mag) variability.

As an initial effort, we acquired IRTF/SpeX spectroscopic data of nine YSOs that show variability with amplitudes of $\simeq 3$ mag at near- to mid-IR wavelengths. We find four YSOs that show similarities to the spectra of known eruptive YSOs. The mid-IR light curves and color-magnitude diagrams of these sources also align with the expected changes influenced by large variations in the accretion rates of these systems.

We attempt to classify these new four outbursts into the known sub-classes of this variability class. We find two objects that can be firmly classified as bonafide FUors, while one YSO shows a mixture of characteristics between FUors and EXors, and is therefore classified as a V1647 Ori-like source. A final YSO has an uncertain outburst duration and can only be temporarily placed as a candidate FUor.

## Acknowledgments

We are grateful to all past and current JKAS authors for their trust and their support.

CCP was supported by the National Research Foundation of Korea (NRF) grant funded by the Korean government (MEST) (No. 2019R1A6A1A10073437). J.-E. Lee was supported by the New Faculty Startup Fund from Seoul National University and the NRF grant funded by the Korean government (MSIT) (grant number 2021R1A2C1011718). ZG is supported by the ANID FONDECYT Postdoctoral program No. 3220029. ZG acknowledges support by ANID, — Millennium Science Initiative Program — NCN19_171. D.J. is supported by NRC Canada and by an NSERC Discovery Grant. PWL received some funding from grant ST/R000905/1 of the UK Science and Technology Facilities Council.

## References

Antoniucci, S., Giannini, T., Li Causi, G., & Lorenzetti, D. 2014, ApJ, 782, 51

Artur de la Villarmois, E., Jørgensen, J. K., Kristensen, L. E., et al. 2019, A&A, 626, A71

Audard, M., Ábrahám, P., Dunham, M. M., et al. 2014, Protostars and Planets VI, 387

Bailer-Jones, C. A. L., Rybizki, J., Fouesneau, M., Demleitner, M., & Andrae, R. 2021, AJ, 161, 147

Baraffe, I., Elbakyan, V. G., Vorobyov, E. I., & Chabrier, G. 2017, A&A, 597, A19

Becker, J. C., Batygin, K., & Adams, F. C. 2021, ApJ, 919, 76

Bell, K. R., & Lin, D. N. C. 1994, ApJ, 427, 987

Boss, A. P. 2013, ApJ, 764, 194

Cieza, L. A., Casassus, S., Tobin, J., et al. 2016, Nature, 535, 258

Cieza, L. A., Ruíz-Rodríguez, D., Perez, S., et al. 2018, MNRAS, 474, 4347

Cleaver, J., Hartmann, L., & Bae, J. 2023, MNRAS, 523, 5522

Connelley, M. S., & Reipurth, B. 2018, ApJ, 861, 145

Contreras Peña, C., Lucas, P. W., Kurtev, R., et al. 2017, MNRAS, 465, 3039

Contreras Peña, C., Herczeg, G. J., Ashraf, M., et al. 2023, MNRAS, 521, 5669

Cruz-Sáenz de Miera, F., Kóspál, Á., Abrahám, P., et al. 2023, A&A, 678, A88

Cushing, M. C., Vacca, W. D., & Rayner, J. T. 2004, PASP, 116, 362

Cutri, R. M., Skrutskie, M. F., van Dyk, S., et al. 2003, VizieR Online Data Catalog, II/246

Dunham, M. M., Stutz, A. M., Allen, L. E., et al. 2014, Protostars and Planets VI, 195

Evans, II, N. J., Dunham, M. M., Jørgensen, J. K., et al. 2009, ApJS, 181, 321

Fischer, W. J., Hillenbrand, L. A., Herczeg, G. J., et al. 2022, arXiv e-prints, arXiv:2203.11257

Gaia Collaboration. 2022, VizieR Online Data Catalog, I/355

Guo, Z., Lucas, P. W., Contreras Peña, C., et al. 2021, MNRAS, 504, 830

Hartmann, L., & Kenyon, S. J. 1996, ARA&A, 34, 207

Herbig, G. H. 1977, ApJ, 217, 693

Herbig, G. H. 1989, in Eur. South. Obs. Conf. Workshop Proc., Vol. 33, 233–246

Herbig, G. H. 2008, AJ, 135, 637

Hillenbrand, L. A., & Rodriguez, A. C. 2022, Res. Notes of the AAS, 6, 6

Hillenbrand, L. A., Contreras Peña, C., Morrell, S., et al. 2018, ApJ, 869, 146

Hubbard, A. 2017, MNRAS, 465, 1910

Kenyon, S. J., Hartmann, L. W., Strom, K. M., & Strom, S. E. 1990, AJ, 99, 869

Kóspál, Á., Cruz-Sáenz de Miera, F., White, J. A., et al. 2021, ApJS, 256, 30

Kryukova, E., Megeath, S. T., Hora, J. L., et al. 2014, AJ, 148, 11

Kuhn, M. A., de Souza, R. S., Krone-Martins, A., et al. 2021, ApJS, 254, 33
Lawrence, A., Warren, S. J., Almaini, O., et al. 2007, MNRAS, 379, 1599
Lee, J.-E., Lee, S., Baek, G., et al. 2019, Nature Ast., 3, 314
Lee, Y.-H., Johnstone, D., Lee, J.-E., et al. 2020, ApJ, 903, 5
Lee, Y.-H., Johnstone, D., Lee, J.-E., et al. 2021, ApJ, 920, 119
Liu, H., Herczeg, G. J., Johnstone, D., et al. 2022, ApJ, 936, 152
Lorenzetti, D., Giannini, T., Larionov, V. M., et al. 2007, ApJ, 665, 1182
Lorenzetti, D., Larionov, V. M., Giannini, T., et al. 2009, ApJ, 693, 1056
Lorenzetti, D., Antoniucci, S., Giannini, T., et al. 2012, ApJ, 749, 188
Lucas, P. W., Hoare, M. G., Longmore, A., et al. 2008, MNRAS, 391, 136
Lucas, P. W., Smith, L. C., Contreras Peña, C., et al. 2017, MNRAS, 472, 2990
Lucas, P. W., Elias, J., Points, S., et al. 2020, MNRAS, 499, 1805
Mainzer, A., Grav, T., Bauer, J., et al. 2011, ApJ, 743, 156
Mainzer, A., Bauer, J., Cutri, R. M., et al. 2014, ApJ, 792, 30
Makin, S. V., & Froebrich, D. 2018, ApJS, 234, 8
Mège, P., Russeil, D., Zavagno, A., et al. 2021, A&A, 646, A74
Minniti, D., Lucas, P., Emerson, J., et al. 2010, New A, 15, 433
Park, W., Lee, J.-E., Contreras Peña, C., et al. 2021, ApJ, 920, 132
Rayner, J. T., Toomey, D. W., Onaka, P. M., et al. 2003, PASP, 115, 362
Reipurth, B., & Aspin, C. 2010, in Evolution of Cosmic Objects through their Physical Activity, ed. H. A. Harutyunian, A. M. Mickaelian, & Y. Terzian, 19–38
Robitaille, T. P., Meade, M. R., Babler, B. L., et al. 2008, AJ, 136, 2413
Saito, R. K., Hempel, M., Minniti, D., et al. 2012, A&A, 537, A107
Saral, G., Hora, J. L., Willis, S. E., et al. 2015, ApJ, 813, 25
Scholz, A., Froebrich, D., & Wood, K. 2013, MNRAS, 430, 2910
Simon, R., Jackson, J. M., Clemens, D. P., Bania, T. M., & Heyer, M. H. 2001, ApJ, 551, 747
Skrutskie, M. F., Cutri, R. M., Stiening, R., et al. 2006, AJ, 131, 1163
Wang, M.-T., Herczeg, G. J., Liu, H.-G., et al. 2023, ApJ, 957, 113
Wang, S., & Chen, X. 2019, ApJ, 877, 116
Wright, E. L., Eisenhardt, P. R. M., Mainzer, A. K., et al. 2010, AJ, 140, 1868

## Appendix A. High-Amplitude YSOs

Here, we present the mid-IR light curves and spectra of five YSOs that were part of the IRTF/SpeX follow-up, but cannot be firmly classified as eruptive YSOs. Amplitudes and classification based on these data are presented in Table A.1. The classification as an evolved source for SPICY 95307 is reached as its spectrum shows $^{13}CO$ absorption bands (see Figure A.2). The latter is characteristic of Asymptotic Giant Branch (AGB) stars (see e.g. Guo et al. 2021).

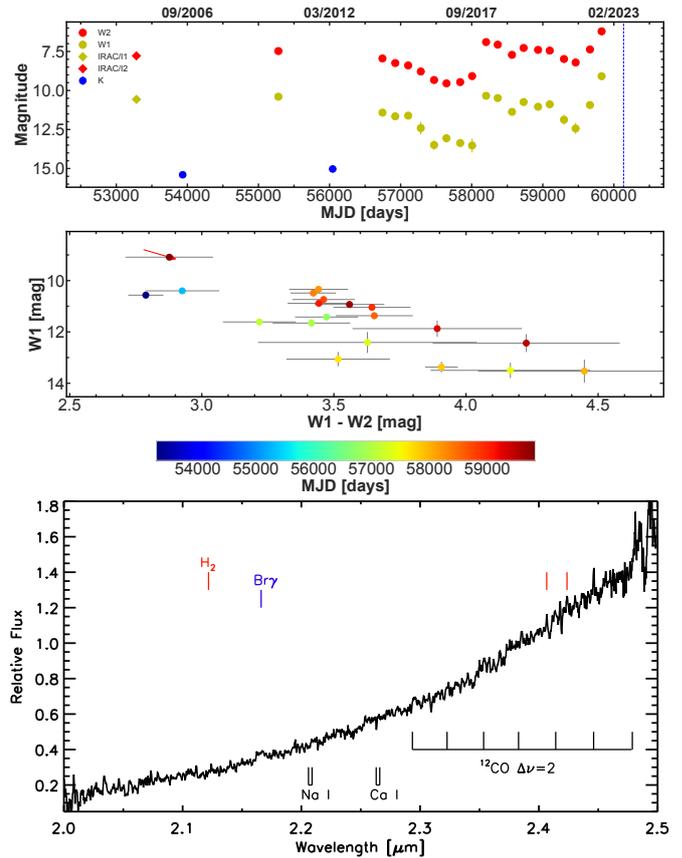

**Figure A.1.** (top) Light curve, (middle) $W1$ versus $W1 - W2$ color-magnitude diagram, and (bottom) IRTF/SpeX spectrum of SPICY 79425. Symbols and labels are the same as in Figure 1.

**Table A.1.** YSO Sample

| YSO ID | Other Name | $\alpha$ (J2000) | $\delta$ (J2000) | Class | $\Delta W1$ | $\Delta W2$ | Spectral Class | Photometric Class | Final Class |
|---|---|---|---|---|---|---|---|---|---|
| SPICY 79425 | WISEA J181725.67−170211.7[a] | 18:17:25.69 | −17:02:11.94 | I | 4.44 | 3.34 | Featureless | V1647 Ori | Candidate |
| SPICY 95397 | WISEA J185720.27+015711.8[b] | 18:57:20.27 | +01:57:12.23 | II | 3.56 | 3.24 | Evolved | long-term | non-YSO |
| SPICY 103300 | 2MASS J19285321+1714565[c] | 19:28:53.24 | +17:14:56.49 | FS | 3.07 | 2.74 | Br$\gamma$ emission | V1647 Ori? | Candidate |
| SPICY 104367 | — | 19:32:26.02 | +19:40:08.85 | I | 2.51 | 3.41 | Noisy | FUor | Candidate |
| SPICY 115884 | MSX6C G081.7203−00.6744[d] | 20:44:18.10 | +41:36:50.34 | I | 4.78 | 3.58 | Featureless? | V1647 Ori? | Candidate |

[a] Previously identified as a high amplitude variable star. Source 15 in Lucas et al. (2020), [b] Previously identified as a high amplitude variable star. Source 21 in Lucas et al. (2020), [c] Classified as a candidate YSO by Robitaille et al. (2008), [d] classified as a protostellar candidate in Cygnus-X by Kryukova et al. (2014).

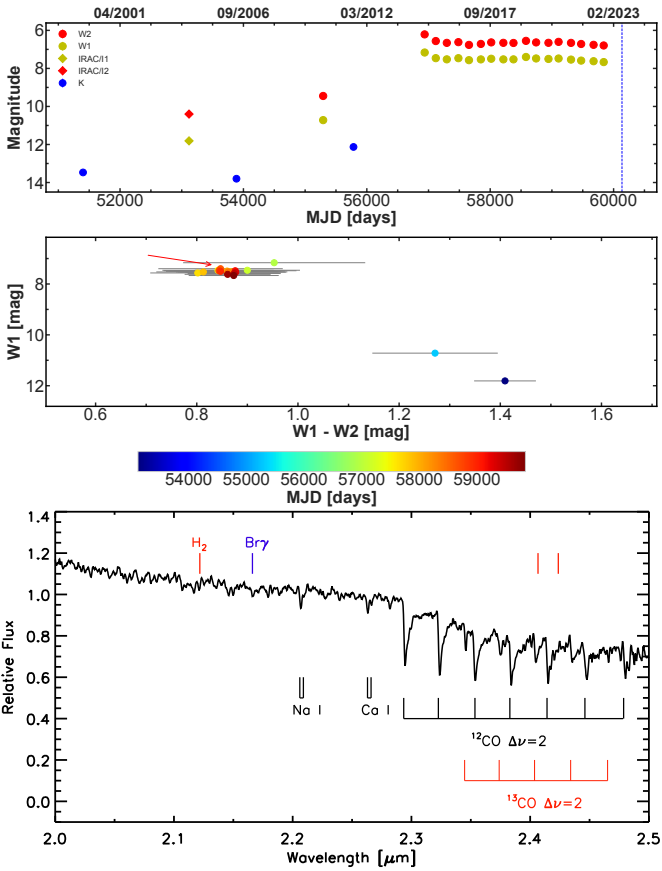

**Figure A.2.** (top) Light curve, (middle) $W1$ versus $W1-W2$ color-magnitude diagram, and (bottom) IRTF/SpeX spectrum of SPICY 95397. Symbols and labels are the same as in Figure 1.

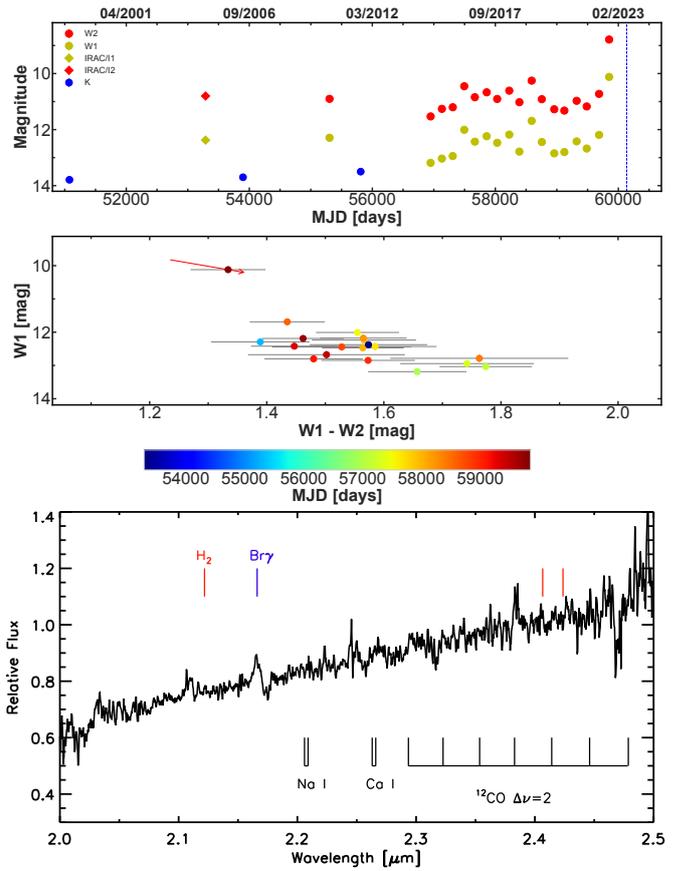

**Figure A.3.** (top) Light curve, (middle) $W1$ versus $W1-W2$ color-magnitude diagram, and (bottom) IRTF/SpeX spectrum of SPICY 103300. Symbols and labels are the same as in Figure 1.

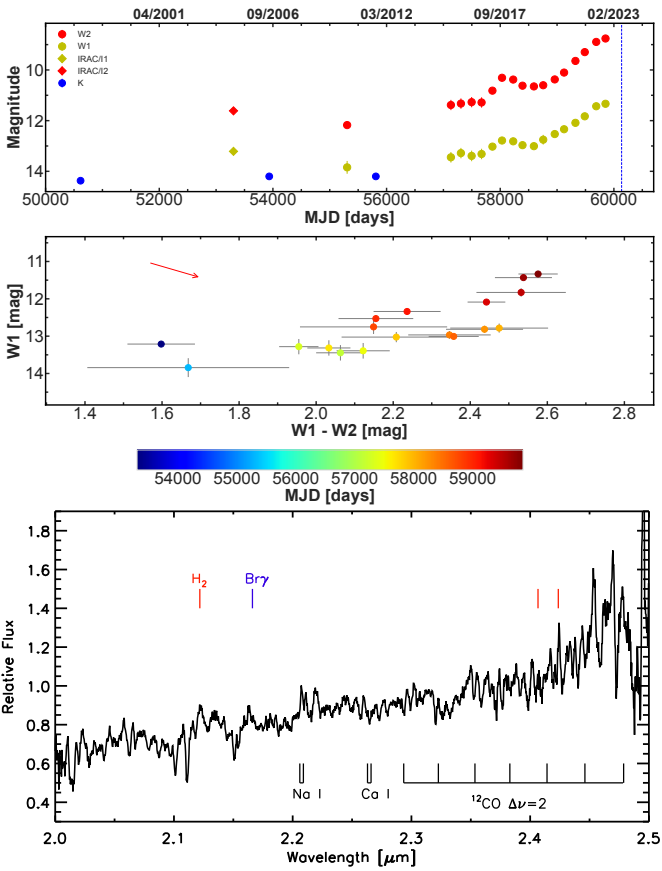
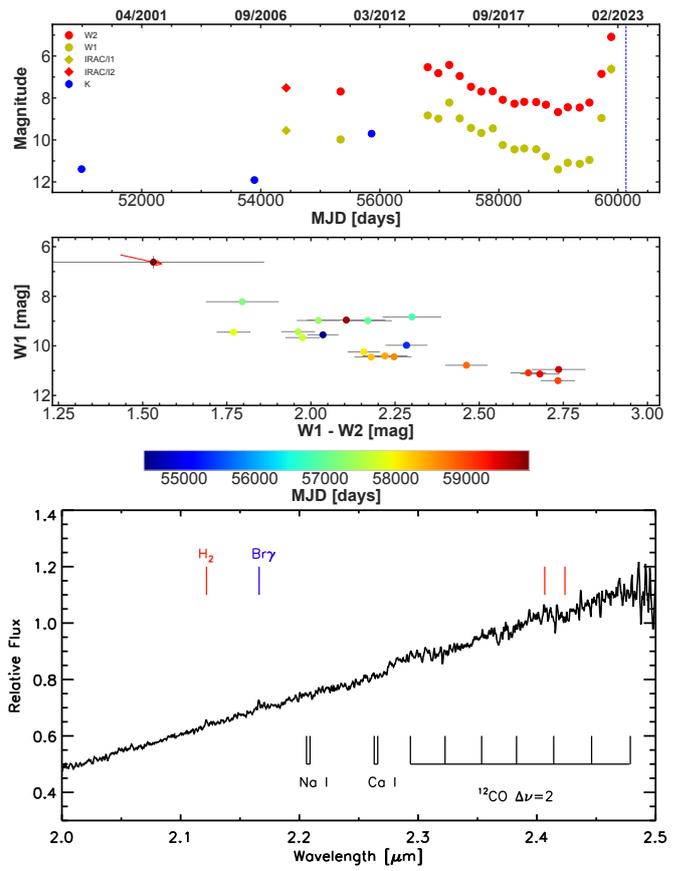

**Figure A.4.** (top) Light curve, (middle) $W1$ versus $W1 - W2$ color-magnitude diagram, and (bottom) IRTF/SpeX spectrum of SPICY 104367. Symbols and labels are the same as in Figure 1.

**Figure A.5.** (top) Light curve, (middle) $W1$ versus $W1 - W2$ color-magnitude diagram, and (bottom) IRTF/SpeX spectrum of SPICY 115884. Symbols and labels are the same as in Figure 1.